\documentclass{ptapap}
\pdfoutput=1
\author{James Barron}[QU, RMC]
\author{Gregg A. Wade}[RMC]
\author{Dominic M. Bowman}[Leuven]
\author{Alexandre David-Uraz}[Del]
\author{Melissa S. Munoz}[QU]
\author{Herbert Pablo}[AAVSO]
\author{Sergio Sim\'on-D\'iaz}[IdA, ULL]

\affil[RMC]{Department of Physics and Space Science, Royal Military College of Canada, PO Box 17000, Kingston, ON, K7K 7B4, Canada}
\affil[QU]{Department of Physics, Engineering Physics \& Astronomy, Queen's University, 64 Bader Lane, Kingston, ON, K7L 3N6, Canada}
\affil[Del]{Department of Physics and Astronomy, University of Delaware, Newark, DE 19716, USA}
\affil[IdA]{Instituto de Astrof\'isica de Canarias,E-38\,200 La Laguna, Tenerife, Spain}
\affil[ULL]{Universidad de La Laguna, Universidad de La Laguna, E-38\,205 La Laguna, Tenerife, Spain}
\affil[Leuven]{Institute of Astronomy, KU Leuven, Celestijnenlaan 200D, B-3001 Leuven, Belgium}
\affil[AAVSO]{AAVSO, 49 Bay State Road, Cambridge, MA 02138, USA}

\title{MOBSTER: Identifying Candidate Magnetic O Stars through Rotational Modulation of \textit{TESS} Photometry}
\headtitle{MOBSTER: Identifying Candidate Magnetic O Stars through \dots}
\begin{document}

\maketitle

\begin{abstract}
Being relatively rare, the properties of magnetic O stars are not fully understood. To date fewer than a dozen of these stars have been confirmed, making any inference of their global properties uncertain due to small number statistics. To better understand these objects it is necessary to increase the known sample. The MOBSTER collaboration aims to do this by identifying candidate magnetic O, B, and A stars from the identification of rotational modulation in high-precision photometry from the Transiting Exoplanet Survey Satellite (\textit{TESS}). Here we discuss the collaboration's efforts to detect rotational modulation in \textit{TESS} targets to identify candidate magnetic O stars for future spectropolarimetric observations.
\end{abstract}

\section{Introduction}
O-type stars are among the most massive and luminous stars with surface temperatures over 30\,000 K. They are the progenitors of black holes and neutron stars and provide chemical enrichment to the surrounding interstellar medium. Unlike low mass stars that generate their magnetic fields through convective dynamos, high-mass stars possess radiative envelopes and lack convection in their outer layers to generate large scale magnetic fields. Nevertheless, spectropolarimetric observations have shown that some O stars possess strong (> 1 kG), oblique, typically dipolar magnetic fields at their surfaces (e.g. \citealp{Martins_HD108, castro_54879, grunhut_mimes}). It is generally hypothesized that these fields are formed at an earlier stage in the star's evolution, and are aptly named \textit{fossil} fields \citep{borra_1982}. The origins of the fossil fields are still debated, whether they are formed during pre-main sequence evolution \citep{fossil_field_pms} or in stellar mergers \citep{stellar_merge}. These surface magnetic fields affect a star's evolution through both magnetic braking \citep{Doula_2009} and interactions with the stellar wind leading to the formation of a magnetosphere \citep{magnetosphere_mhd}.

To date there are only 11 confirmed magnetic O stars \citep{wade_2015_ostars}, giving a magnetic incidence rate of less than 10\%  \citep{grunhut_mimes}. The small number of stars available for study makes it difficult to model their evolution and infer global properties. It is necessary to identify more magnetic O stars to increase our knowledge about their origin and evolutionary paths. The Transiting Exoplanet Survey Satellite (\textit{TESS}; \citealp{ricker}) offers an unparalleled opportunity to study stellar variability across the HR diagram due to its precision, high cadence and comprehensive coverage of the sky ($\sim$85\%). The MOBSTER collaboration (Magnetic OB[A] Stars with \textit{TESS}: probing their Evolutionary and Rotational
properties; \citealp{mobster1, mobster2, mobster3}) aims to make use of \textit{TESS} data to search for rotational modulation in OBA stars to identify magnetic massive star candidates for follow-up spectropolarimetric observations. For further discussion about the collaboration and rotational modulation as seen in B and A-type stars see \cite{uraz_vienna} and \cite{uraz_london}. Here we discuss our efforts towards identifying magnetic O star candidates through rotational modulation in \textit{TESS} photometry.

\section{Magnetospheres and Rotational Modulation}
In OB stars that possess surface magnetic fields, the radiatively-driven stellar wind is confined by the closed field lines around the magnetic equator, forming a magnetosphere. As the star rotates, the column density along the line-of-sight changes due to the misalignment between the rotational and magnetic axes. Therefore the continuum intensity of light varies during the rotation phase, leading to periodic variations in the star's photometric light curve (e.g. \citealp{munoz_ADM}). Rotational modulation is also seen in later B and A type stars. In this case the magnetic fields influence the atomic diffusion process in the star's atmosphere \citep{alecian}, leading to chemical inhomegeneities on the stellar surface which are seen as brightness spots. Such variations have been successfully detected in known magnetic B stars \citep{mobster1} using a Lomb-Scargle (L-S) analysis (see \citealt{lomb_scargle} for an introduction to the technique). The rotational signature typically manifests itself in the periodogram as a peak at the rotation frequency and a number of harmonics, which depend on the spatial distribution of spots on the stellar surface with respect to the observer (e.g. \citealp{Stibbs1950, bowman_rotation, mobster2}).

However, such a signature is not unique to rotational modulation, and can also be associated with ellipsoidal variations in binary systems. Before \textit{TESS}, high precision space photometric surveys of O stars, while limited, have found diverse types of variability including periodic variations, either due to rotation or binarity and stochastic low-frequency signals (e.g. \citealp{Blomme2011b, Buysschaert}). These stochastic signals have been attributed to both internal gravity waves (e.g. \citealp{Aerts_IGW, Bowman2019b, bowman_IGW}) and subsurface convection zones \citep{subsurface_convection}. A study of the variability of O and B stars using 2-min \textit{TESS} data for sectors 1 and 2 has been conducted by \cite{pedersen} and contained 5 O star targets, of which 3 were found to exhibit rotational modulation. 

To aid in the diagnosis of rotational modulation, other criteria must be considered. Literature searches and examination of available spectra can help diagnose variability due to stellar companions. The projected rotational velocity found from spectroscopy provides a lower limit for the star's rotational frequency, and an upper limit can be placed by the star's critical rotation velocity. Light curves that are found to show rotational modulation can then be modelled to constrain parameters relating to the star's magnetic field to help determine the feasibility of spectropolarimetric follow-up observations. The Analytic Dynamical Magnetosphere (ADM) model developed by \cite{ADM} offers a time-averaged description of the density and velocity structure of the magnetosphere of slowly rotating magnetic massive stars. The model is in good agreement with magnetohydrodynamic simulations and can be used to determine magnetic and stellar parameters from photometry \citep{munoz_ADM}. The existence of rotational modulation alone does not necessarily imply the existence of a magnetosphere, but it can be used as a diagnostic tool to identify candidate magnetic O-type stars.

\section{\textit{TESS} Observations}
 The \textit{TESS} survey divides the sky into the north and south ecliptic hemispheres, each containing 13 partially overlapping sectors that are observed nearly continuously for 27 days. Target pixel files (TPFs) for 200\,000 pre-selected targets are available at 2-min cadence, and full-frame images (FFIs) are available at a 30-min cadence for approximately 470 million point sources \citep{stassun}. In addition to the TPFs, 2-min processed (\texttt{PDCSAP}) light curves are provided by the \textit{TESS} Science Team \citep{TESS_data} through the Mikulski Archive for Space Telescopes (MAST). It is important to note that \textit{TESS} pixels are relatively large (21 $\times$ 21 arcsec), and so targets may include flux from multiple nearby stars (e.g. Fig \ref{fig:TR1622}). Light curve extraction from \textit{TESS} FFIs can be done using open-source tools such as \texttt{eleanor} \citep{eleanor}. The tool can perform background subtraction, instrument systematics decorrelation, principal component analysis, and point-spread function modelling. As of July of 2019 observations of the southern ecliptic were concluded, allowing for the opportunity to perform a comprehensive search of southern O star targets for rotational modulation.
 
\section{Known Magnetic O Stars}
To date 9 of the 11 confirmed magnetic O-type stars have been observed by \textit{TESS}. Of the sample HD\,148937 ($P_{\rm rot} = $ 7.03 d), HD\,47129 (1.21 d), HD\,37742 (7.0 d) have rotational periods that are sufficiently short for them to have been observed for more than a full rotational cycle \citep{petit_2013, grunhut_mimes}. All 9 magnetic stars show variability on timescales of 1-6 days and analysis is ongoing to determine the source(s) of the variability. Here we discuss our preliminary findings from the analysis of the magnetic O stars, which demonstrate the need for care in working with \textit{TESS} photometry.

\begin{figure}[!ht]
\includegraphics[width=\textwidth]{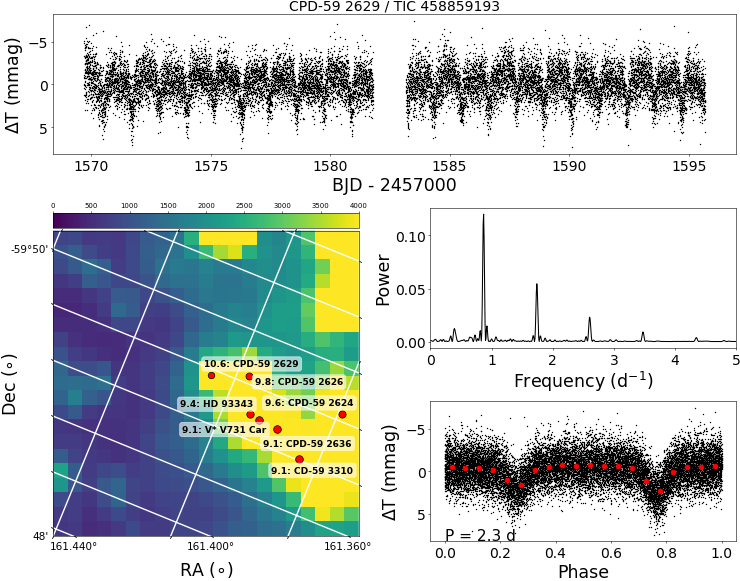}
\caption{\textbf{Top}: \textit{TESS} light curve of CPD-59 2629 (Tr 16-22), sector 10, 2-min \texttt{PDCSAP} flux. Variability does not appear to be from CPD-59 2629 but the nearby eclipsing binary V731 Car. \textbf{Left}: FFI pixel cutout centered on target. Stars shown are from crossmatch with Gaia DR2 in 200 arcsecond radius for apparent $g$-band magnitude less than 11. Numbers denote $g$-band magnitude and names come from crossmatch with SIMBAD using \texttt{astroquery}. The colour scale denotes flux. \textbf{Top Right}: L-S periodogram of light curve. \textbf{Bottom Right:} Light curve phased on $P$ = 2.3 days, which corresponds to the known binary period of V731 Car. Red circles denote binned points.}
\label{fig:TR1622}
\end{figure}

Figure \ref{fig:TR1622} shows the 2-min \texttt{PDCSAP} light curve of CPD-59 2629 (Tr 16-22), a magnetic O8.5V star with a known rotation period of approximately 54 days \citep{TR1622_period}. However, the principal variability in the extracted \textit{TESS} light curve appears to be due to a blended eclipsing binary. The peak of maximum power in L-S periodogram corresponds to a frequency of 0.87-d$^{-1}$. A pixel cutout of a sample FFI image containing CPD-59 2629 (apparent $g$-band magnitude of 10.6) is overlayed with all stars ($g < 11$) in a 200 arcsec radius from a crossmatch with the Gaia DR2 database using \texttt{astroquery}. A literature search shows that V731 Car ($g = 9.1$), approximately 90 arcsec (4 pixels) away is an eclipsing binary with a period of 2.3 days \citep{V731_Car}. This corresponds to the peak at 0.43~d$^{-1}$ detected in the periodogram, and the light curve is shown phased to this period in the bottom right panel of Figure 1. In fact, all stars shown on this FFI cutout are O-type stars and may be at least partially contaminated by this eclipsing binary system. This example demonstrates the necessity of checking bright sources near a given target to assess that any variability seen is real and not due to contamination.

\begin{figure}[!ht]
\includegraphics[width=\textwidth]{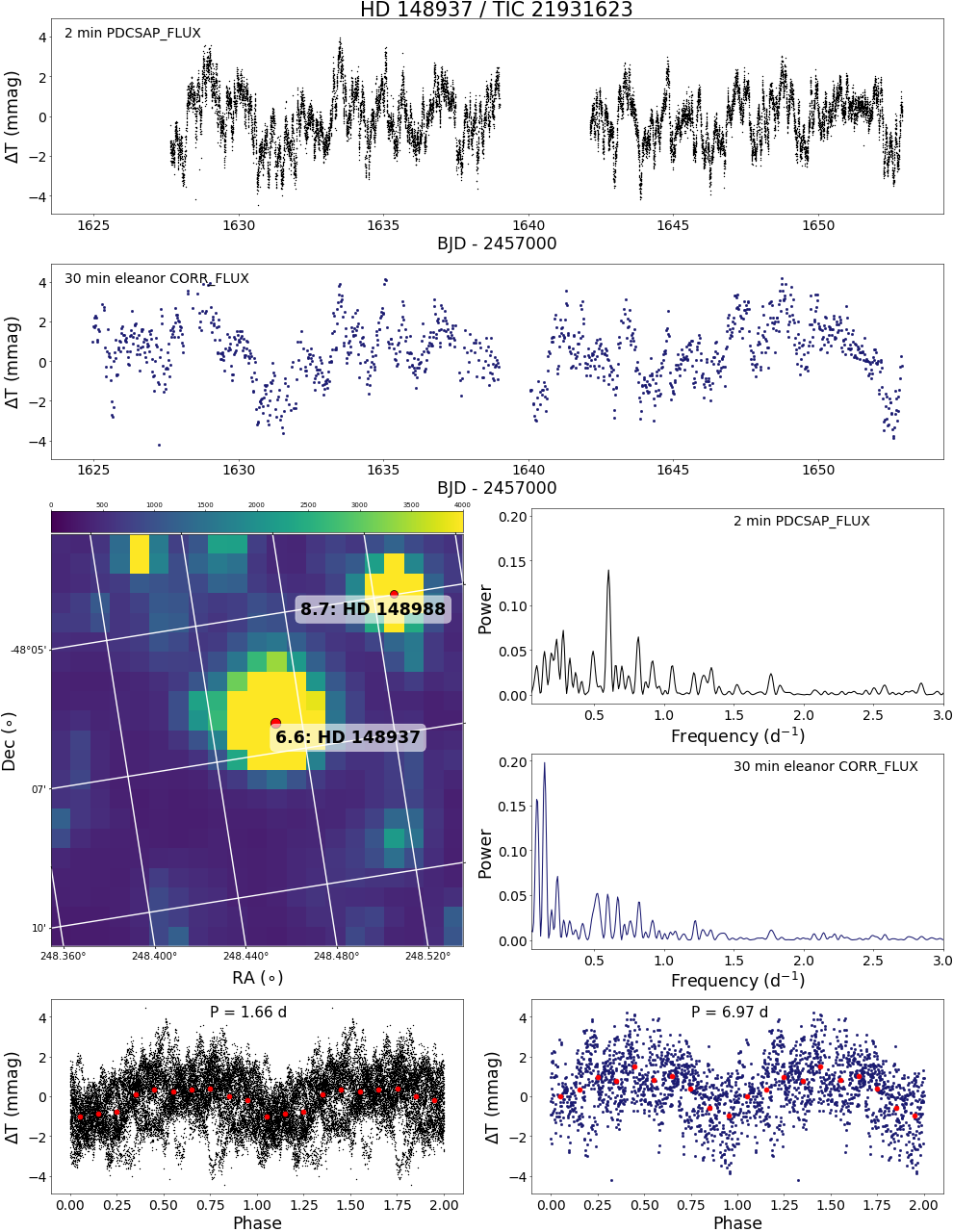}
\caption{\textbf{Top}: \textit{TESS} Light curve of HD\,148937 from 2-min cadence data, sector 12 (black). Below is the light curve extracted from 30-min FFI using \texttt{eleanor} (blue). \textbf{Left}: Same as Fig. \ref{fig:TR1622} except for HD\,148937. \textbf{Right}: L-S periodograms of both light curves. \textbf{Bottom}: Light curves phased to dominant periods, $P = 1.66$ d (2-min), $P = 6.97$ d (30-min).}
\label{fig:HD148937}
\end{figure}

In the analysis of HD\,148937 (Fig. \ref{fig:HD148937}) we find a discrepancy between the 2-min processed light curve, and 30-min FFI extracted light curve obtained with \texttt{eleanor}. Figure \ref{fig:HD148937} shows the 2-min light curve for HD\,148937, a magnetic O6f?p type star with a 7.03-day rotation period \citep{HD148937_period}. The L-S periodogram of the 2-min cadence data returns a period of $P = 1.66$ days, and the data are shown phased to this period in the bottom left. Also shown is the 30-min \texttt{eleanor} FFI extracted \texttt{CORR\_FLUX} light curve. It is generated from the raw FFI flux from an aperture centered on the target and corrected for systematic effects. The L-S periodogram returns a 6.97 day period, which is close to the published value of the rotational period (7.03 d). The 2-min light curve is missing data from the start of the observing window and has a larger section removed in the middle. Other sources of discrepancy between the two methods may include the detrending of \textit{TESS} systematics, as well as the difference in size and shape of the apertures. This shows that comparisons between 2-min cadence and 30-min FFI extracted light curves may be necessary to check for consistency and to identify the true rotation period of magnetic stars.

\section{Search for Candidate Rotational Variables}
\begin{figure}[!ht]
\includegraphics[width=\textwidth]{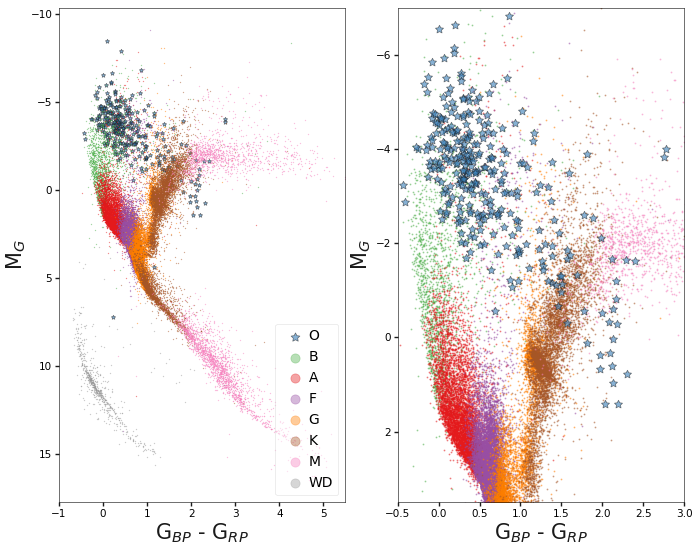}
\caption{Gaia colour-magnitude diagram of southern 2-min \textit{TESS} targets crossmatched with Gaia DR2 database using MAST (approx. 47\,000 stars). Overlayed is the southern GOSC sample (315). Apparent $g$-band magnitudes are converted to absolute $G$-band magnitudes using distances from \cite{bailer_jones}. Spectral types other than O stars are colour coded according to listed spectral type in SIMBAD.}
\label{fig:cm}
\end{figure}
Our goal is to perform a comprehensive search for rotational modulation in O stars in the southern ecliptic (sectors 1-13), utilizing both 2-min cadence processed light curves and 30-min FFI extraction. The Galactic O-Star Catalog (GOSC; \citealp{gosc}) provides spectral classifications of bright O-type stars and will serve as the primary source for our sample. Other sample selection methods were considered including crossmatching targets with SIMBAD, however, we have found that SIMBAD may list old and inaccurate spectral classifications. This is especially true for O stars, which often require high-resolution spectroscopy such as from the IACOB and OWN surveys \citep{iacob_ref, own_survey}. Using Gaia DR2 photometry \citep{gaiaDR2} we have placed our sample of southern GOSC stars on a colour-magnitude diagram of all southern 2-min cadence targets crossmatched from the Gaia DR2 database (Fig. \ref{fig:cm}). The apparent $g$-band magnitudes are converted to absolute magnitudes using distances determined by \cite{bailer_jones} (approx. 47\,000 stars). We note that we have not performed any corrections for reddening or extinction. All targets are colour-coded according to their listed spectral type in SIMBAD except for the O-stars. Blue stars denote all southern GOSC targets that also have distance determinations from \cite{bailer_jones} (315 targets). It is apparent from these diagrams that it would be difficult to perform a sample selection of O stars using a colour-magnitude selection.
\begin{figure}[!ht]
\includegraphics[width=\textwidth]{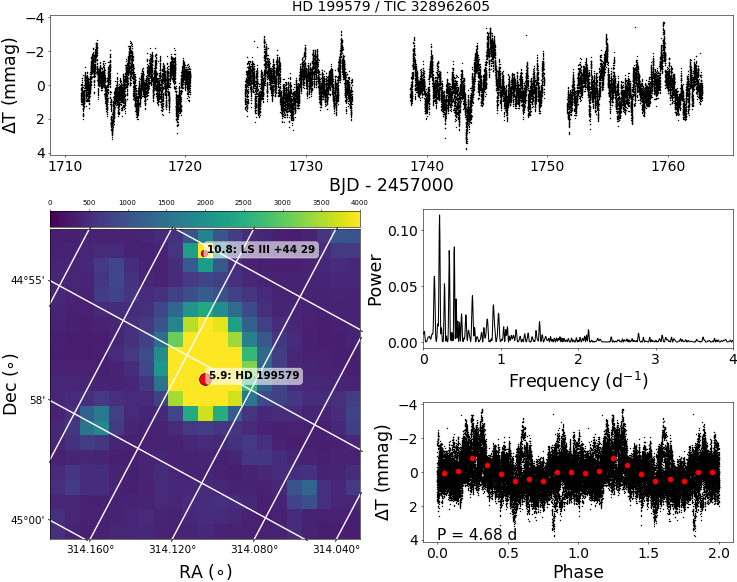}
\caption{Same as Fig. \ref{fig:TR1622} for HD\,199579. The frequency of max power corresponds to a period of 4.68 days. The next 4 most significant peaks found from a pre-whitening procedure correspond to periods between 1.1 and 2.9 days.}
\label{fig:HD199579_4plot}
\end{figure}

Figure \ref{fig:HD199579_4plot} shows an example target (HD\,199579) that we have flagged as potentially showing rotational modulation with a period of 4.68 days. This star had previously been determined to be a candidate magnetic star by \cite{grunhut_mimes}, and may warrant follow up spectropolarimetric observations. It also appears to be multiperiodic and show stochastic low-frequency variability similar to other OB stars \citep{Bowman2019b}.

\section{Next Steps}
Further work is required to confirm the suitability of the 2-min processed \textit{TESS} light curves in the search for O star rotational modulation. Comparisons to light curves obtained by other FFI extraction methods should be made to determine the best method and thus inform subsequent detrending techniques for the removal of systematics. Possible contamination due to crowded regions will be checked using a \texttt{python} tool we have developed to generate the pixel images seen in Figs. \ref{fig:TR1622}, \ref{fig:HD148937} and \ref{fig:HD199579_4plot}. We will perform a Lomb-Scargle analysis on all southern O star targets in our GOSC sample. A large number of GOSC stars have spectra available from the IACOB and OWN surveys, and have been analyzed to determine stellar parameters (e.g. \citealp{IACOB}). Light curves that are determined to show variability due to rotational modulation can be modelled using ADM to determine magnetic field geometry and strength and used to judge the feasibility of detection in spectropolarimetric observations. Analysis of these observations and any magnetic field detections will provide valuable insight into the photometric signature of rotationally modulated O stars and will refine our method for future studies.

\acknowledgements{GAW acknowledges Discovery Grant support from the Natural Sciences and Engineering Research Council (NSERC) of Canada. ADU gratefully acknowledges the support of the Natural Science and Engineering Research Council of Canada (NSERC). The research leading to these results has received funding from the European Research Council (ERC) under the European Union's Horizon 2020 research and innovation programme (grant agreement No. 670519: MAMSIE). S-SD acknowledges support from the Spanish Government Ministerio de Ciencia, Innovaci\'on y Universidades through grant PGC-2018-091\,3741-B-C22. This research includes data collected with the TESS mission, obtained from the MAST data archive at the Space Telescope Science Institute (STScI). Funding for the TESS mission is provided by the NASA Explorer Program. STScI is operated by the Association of Universities for Research in Astronomy, Inc., under NASA contract NAS 5–26555. This research has made use of the SIMBAD database, operated at CDS, Strasbourg, France.}

\bibliographystyle{ptapap}
\bibliography{Barron}

\end{document}